\documentclass[comment,prl,twocolumn,nofootinbib, preprintnumbers, superscriptaddress]{revtex4}

\usepackage{amsmath,amssymb,bm,slashed,braket}
\usepackage{graphicx}
\usepackage{epstopdf}
\usepackage{float}
\usepackage[colorlinks=true,
            linkcolor=blue,
            urlcolor=blue,
            citecolor=green,          
            bookmarks=true,
            bookmarksnumbered=true,
            breaklinks=true,
            pdfpagemode=Fullscreen,
            pdfstartview=FitBH]{hyperref}

\usepackage[normalem]{ulem}
\usepackage{color}

\definecolor{Orange}{cmyk}{0,0.61,0.87,0}
\definecolor{JungleGreen}{cmyk}{0.99,0,0.52,0}
\definecolor{OliveGreen}{cmyk}{0.64,0,0.95,0.40}
\definecolor{Brown}{cmyk}{0,0.81,1,0.60}
\definecolor{RoyalBlue}{cmyk}{0.71,0.53,0,0.12}
\definecolor{Gray}{cmyk}{0,0,0,0.40}
\definecolor{LightPink}{cmyk}{0.0,0.25,0,0}
\definecolor{LLightPink}{cmyk}{0.0,0.10,0,0}
\definecolor{LightBlue}{cmyk}{0.25,0,0,0}
\definecolor{LightGray}{cmyk}{0,0,0,0.2}

\usepackage{xcolor}
\definecolor{gesfpurple}{rgb}{0.47,0.19,0.42}

\definecolor{gesflanse}{rgb}{0.00,0.50,0.50}

\definecolor{gesfblue}{rgb}{0.08,0.42,0.76}

\definecolor{gesfred}{rgb}{1,0,0}

\definecolor{gesfwhite}{rgb}{1,1,1}

\definecolor{gesfblack}{rgb}{0,0,0}

\newcommand{\geqn}[1]{Eq.\,\hypersetup{linkcolor=blue}(\ref{#1})\hypersetup{linkcolor=blue}}
\newcommand{\gfig}[1]{{\hypersetup{linkcolor=violet}Fig.\,\ref{#1}\hypersetup{linkcolor=blue}}}

\graphicspath{{figs/}}

\begin{document}

\title{
Degeneracy Enhancement of Neutron Oscillation in Neutron Star
}

\author{Xuan-Ye Fu}
\email{fxyinx3196@sjtu.edu.cn}
\affiliation{Zhiyuan College, Shanghai Jiao Tong University, Shanghai 200240, China}

\author{Shao-Feng Ge}
\email[Corresponding author: ]{gesf@sjtu.edu.cn}
\affiliation{Tsung-Dao Lee Institute \& School of Physics and Astronomy, Shanghai Jiao Tong University, China}
\affiliation{Key Laboratory for Particle Astrophysics and Cosmology (MOE) \& Shanghai Key Laboratory for Particle Physics and Cosmology, Shanghai Jiao Tong University, Shanghai 200240, China}

\author{Zi-Yang Guo}
\email{waibiwaibigzy@sjtu.edu.cn}
\affiliation{School of Physics and Astronomy, Shanghai Jiao Tong University, Shanghai 200240, China}

\author{Qi-Heng Wang}
\email{wangq1h@sjtu.edu.cn}
\affiliation{Zhiyuan College, Shanghai Jiao Tong University, Shanghai 200240, China}

\begin{abstract}
We explore the fermion oscillation in a degenerate
environment. The direct consequence is introducing
a Pauli blocking factor $1 - f_i$, where $f_i$ is
the phase space distribution function, for each
intermediate mass eigenstate during propagation. It is then
much easier for a state with larger existing
fraction or density to oscillate into other states
with less degeneracy while the reversed process
is not enhanced. This can
significantly modify the oscillation behaviors.
We apply this degenerate fermion oscillation to
a concrete scenario of neutron-antineutron oscillation
in neutron star. It turns out antineutrons
receive a standing fraction to
annihilate with the environmental neutrons.
The subsequent neutron star heating can put an
extremely stringent bound on the baryon number
violating cross mass term between neutron and antineutron.
The neutron-mirror-neutron oscillation in neutron star
with cooling effect should have similar results.
\end{abstract}

\maketitle 

{\bf Introduction} --
The Grand Unification Theory (GUT)
\cite{Georgi:1974sy,Pati:1974yy,Mohapatra:1974hk,Fritzsch:1974nn} is a fundamental
theory that seeks to unify the electromagnetic, weak,
and strong forces \cite{Croon:2019kpe}.
In addition to proton decay,
the neutron-antineutron ($n \bar n$) oscillation is also
an important prediction of some GUT models \cite{Mohapatra:1996pu}.
While proton decay violates the baryon number by
one, $n \bar n$ oscillation appears to be a
higher-order process with baryon number violation by two.
These could be an important
source for the baryon asymmetry observed in our Universe
\cite{Fukugita:2002hu,Bodeker:2020ghk}.

Previous experiments on $n \bar n$ oscillation have
primarily focused on the processes happening inside atomic nuclei
\cite{Kamiokande:1986pyk,Chung:2002fx,Friedman:2008es} or
in a vacuum environment \cite{Baldo-Ceolin:1994hzw}.
These experiments have provided valuable constraints on
the oscillation parameters, in particular a lower limit
on the oscillation time scale
$\tau_{n\bar n} \sim 1 / \delta m$, where $\delta m$
is the off-diagonal mass matrix element that stands
for neutron-antineutron mixing, up to
the level of $10^8$ seconds and hence a lower limit
$10^{4.5}$\,GeV for the GUT scale \cite{Buccella:1999wy,Babu:2012vc}.
In comparison,
the proton decay sensitivity can reach $10^{34}$ years
for lifetime and $10^{14}$\,GeV for the GUT scale
\cite{Dev:2022jbf,Ohlsson:2023ddi}.

The neutron star (NS) has been used as a natural laboratory for
studying exotic new physics phenomena \cite{Raffelt:1996wa}.
It is also possible to use NS for probing the
$n \bar n$ oscillation \cite{Krishan:ApSS82,Buccella:1987pv}.
However, the previous studies show that the
$n \bar n$ oscillation would not significantly
impact the NS evolution \cite{Buccella:1987pv,Buccella:1999wy}.
In comparison with the terrestrial experiments that
can detect individual annihilation signal induced by the
$n \bar n$ oscillation, the NS heating can only
be sensitive to the macroscopic effect. The NS constraint
ceased from study and was only briefly commented by
\cite{McKeen:2021jbh} and almost touched by \cite{Berezhiani:2020vbe}.

In this paper, we first point out the degenerate scenario
of fermion oscillation in a general context and then apply
it to the $n \bar n$ oscillation inside NS. Since the
neutron matter inside NS is highly degenerate, it is much
easier for neutron to oscillate into antineutron
but the reverse process is much more difficult.
This enhancement of $n \bar n$ oscillation is very significant and allows NS to provide
much stronger constraint than previously expected.

\vspace{1em}

{\bf Fermion Oscillation with Degeneracy} --
The formalism for fermion oscillation in vacuum is well established
\cite{Pontecorvo:1957cp,MNS,P67}.
The initial flavor state $|\alpha \rangle$ can be decomposed
into a linear combination of mass eigenstates $|i\rangle$,
$| \alpha \rangle = \sum_i U^*_{\alpha i} |i\rangle$
with $U_{\alpha i}$ being the mixing matrix element
\cite{Kayser:2004wmw}.
With definite dispersion relation and hence propagating as
plane wave, the mass eigenstate evolves
independently, $| i(t,L) \rangle = e^{i \phi_i(t, L)} |i\rangle$
where $\phi_i(t, L)$ is the corresponding evolution
phase over time interval $t$ and distance $L$.
Correspondingly, the original flavor state
$|\alpha (t, L) \rangle$ also evolves
with time and distance.
For measurement also in a flavor state $|\beta \rangle$,
the transition probability $P_{\alpha \beta}$
is no longer simply $\delta_{\alpha \beta}$ but
\begin{align}
  P_{\beta \alpha}
=
\left|
  \sum_i 
  U_{\beta i} e^{i \phi_i(t, L)} U^*_{\alpha i}
\right|^2.
\label{eq:Aal}
\end{align}
With different evolution phases $\phi_i \neq \phi_j$,
their interference leads to oscillatory behavior in the
transition probability $P_{\alpha \beta}$.
For neutrino oscillation, the individual evolution phase is
$\phi_i \equiv m^2_i L / 2 E_\nu$ and the oscillation phase
is $\Delta \phi_{ij} = \Delta m^2_{ij} L / 2 E_\nu$
with $\Delta m^2_{ij} \equiv m^2_i - m^2_j$.
Note that our notations in \geqn{eq:Aal} are kept general
to cover both neutrino and other fermion oscillations.

The vacuum fermion oscillation is established by considering
fermion propagation on the vacuum state. However, if the
fermion propagates through a degenerate environment fully
filled with fermions of the same type, the amplitude into a
certain mass eigenstate can be suppressed due to the Pauli
exclusion principle which will significantly affect the
oscillation probabilities. In a degenerate environment,
the vacuum state $|0\rangle$ should be replaced by the
degenerate environment state $|\Omega\rangle$.

The fermion in propagation is
created by a creation operator $a^\dagger_{\bm p i}$
and then propagate in the dense environment before
being annihilated by $a_{\bm p i}$. The corresponding
amplitude takes the form as
$\langle \Omega| a_{\bm p i} a^\dagger_{\bm p i} |\Omega \rangle$. For
simplicity, we have omitted other parts such as the
production and detection vertices. Note
that the creation and annihilation operators are always
associated with a physical mass eigenstate that has
definite dispersion relation. If the mass eigenstate
propagate over distance, the evolution phase
$e^{i \phi_i (t, L)}$ should also be taken into
consideration.

The product of creation and annihilation operators
is closely related to the particle momentum density,
$ \hat n_{\boldsymbol p i}
\equiv
  a^\dagger_{\boldsymbol p i} a_{\boldsymbol p i}$.
When applying to the environmental state $|\Omega \rangle$,
the number density operator $\hat n_{\bm p i}$ should
produce the particle momentum density,
$\hat n_{\bm p i} | \Omega \rangle = n_{\bm p i} | \Omega \rangle$ with
$n_{\bm p i} \equiv \int \mathrm d^3 \bm x f_i(\bm x , \bm p)$
%
where $f_i(\bm x, \bm p)$ is the corresponding phase
space distribution function.
For a fermion, the creation and annihilation operators
anticommute as,
  $\{
  a_{\boldsymbol p_1 i},
  a_{\boldsymbol p_2 i}^\dagger 
   \}
=
  (2 \pi)^3 
  \delta^{(3)} (\boldsymbol p_2 - \boldsymbol p_1)$
\footnote{During revision, we notice the same $\delta^{(3)}$ function
has also been obtained in \cite{Ghosh:2022nzo,Barbosa:2024tty}.}.
Then the product $a_{\bm p i} a^\dagger_{\bm p i}$ gives
the Pauli blocking factor \cite{Barbosa:2024tty},
\begin{align}
  \displaystyle{a_{\boldsymbol p i}
  a_{\boldsymbol p i}^\dagger | \Omega \rangle
=
  \int \mathrm d^3 \bm x [1 -f_i(\boldsymbol x , \boldsymbol p)]} | \Omega \rangle,
\end{align}
where the $\delta^{(3)} (\boldsymbol p_2 - \boldsymbol p_1)$
has been removed by momentum integration in the second
quantized fermion field operator,
$\psi(x) \sim \int \mathrm d^4 p (a u e^{- i p \cdot x}
+ b^\dagger v e^{i p \cdot x})$.
  
The scattering amplitude for the whole process composed
of production, propagation, and detection takes the form as,
\begin{align}
\hspace{-3mm}
  \mathcal M_{\beta\alpha}
& \equiv
  \mathcal M_d
\left[
  \sum_i U_{\beta i} U^*_{\alpha i}
  e^{i p_i \cdot (x - y)}
  \langle \Omega |
  a_{\bm p i} a^\dagger_{\bm p i}
  | \Omega \rangle
\right]
  \mathcal M_p,
\label{eq:M}
\end{align}

The fermion oscillation process is a very special
type with the intermediate particle being on shell
and the particle momentum does not change.
These intermediate particles remain
in coherence and evolve according to definite dispersion
relation. The creation and annihilation operators from
the intermediate fields applies on the
thermal vacuum $|\Omega \rangle$ to produce the Pauli
exclusion factor,
$\langle \Omega | a_{\bm p i} a^\dagger_{\bm p i} | \Omega \rangle = 1 - f_i$.
The two on-shell spinors $\bar u$ and $u$
have been absorpted into the production and detection
matrix elements $\mathcal M_p$ and $\mathcal M_d$,
respectively.

Attaching a Pauli blocking factor $1 - f_i$ with an
on-shell intermediate particle is fully consistent
with the thermal quantum field theory (QFT) \cite{Lundberg:2020mwu}.
The fermion propagator in an ensemble of particles,
$S_F(p)
\propto
  i / (p^2 - m^2 + i \epsilon)
- 2 \pi \delta (p^2 - m^2)
  \Theta (p^0) f(\bm p)$
\cite{Ghosh:2022nzo,Notzold:1987ik}
would reduce to
\begin{align}
  S_F
& \propto
  \int \frac {d^3 \bm p}{(2\pi)^3}
  \frac 1 {2 E_{\bm p}}
  e^{-ip\cdot(x-y)}
  [1-f(\bm p)]
  \left(\slashed p + m_\nu \right),  
\end{align}
after Fourier transformation $\int d^4 p / (2 \pi)^4$.
The first term in the
square bracket comes from the ordinary propagator
$i / (p^2 - m^2 + i \epsilon)$ by extracting the
physical pole while the $f(\bm p)$ term from
the thermal contribution after integrating away
the on-shell $\delta(p^2 - m^2)$ function.

Treating the $1 - f$ factor as field strength renormalization,
our result can
be readily obtained with the quantum kinetics theory
\cite{Sigl:1993ctk,Volpe:2023met}. From the mass
matrix diagonalization, the flavor fields $\psi_\alpha$
connect with the mass eigen-fields $\psi_i$ via
a unitarity mixing matrix $U$,
$\psi_\alpha = \sum_i U_{\alpha i} \psi_i$.
By field redefinition, $\psi_i = \sqrt{Z_i} \psi'_i$
with $Z_i \equiv 1 - f_i$,
the canonical field $\psi'_i$ develops evolution phase
$e^{i \phi_i(t, L)}$. Then the scattering matrix
element should contain a factor of
$U_{\beta i} (1 - f_i) e^{i \phi_i} U^*_{\alpha i}$
in the flavor basis,
which is consistent with \geqn{eq:M}.

In the QFT language, the detected
particle number of flavor $\beta$ is proportional to
$N_\beta \propto |\mathcal M_{\beta \alpha}|^2$. To be more
specific, the total particle number without propagation
($x = y$) is proportional to
$N \propto \sum_\beta |\mathcal M_{\beta \alpha}(x = y)|^2$.
After propagating over a nonzero space-time interval $x - y$,
the detected particle number of flavor $\beta$ is proportional
to $N_\beta(x - y) \propto |\mathcal M_{\beta \alpha}(x - y)|^2$.
In other words, a fraction of $N_\beta(x - y) / N$ has
oscillated to flavor $\beta$ which gives the probability for
a neutron to be probed as an antineutron,
\begin{align}
  P_{\alpha \beta}
\equiv
  \frac{
  \left| \sum_i U_{\beta i} U^*_{\alpha i}
  e^{i p_i \cdot (x - y)}
  (1 - f_i) \right|^2
  }{
  \sum_\beta
 \left| \sum_i U_{\beta i} U^*_{\alpha i}
  e^{i p_i \cdot (x - y)}
  (1 - f_i) \right|^2
  }.
\end{align}
With an empty environment
($f_i = 0$), the transition probability reduces to the vacuum
case as shown in \geqn{eq:Aal}.
The same formalism of degenerate oscillation should
also apply for bosons with the Pauli blocking factor $1 - f_i$
replaced by the Bose enhancement factor $1 + f_i$.

\vspace{1em}

{\bf Neutron-Antineutron Oscillation with Degeneracy} --
Being neutral particles, neutron and antineutron
can have cross mass term \cite{Phillips:2014fgb}.
To be exact, the neutron and antineutron Hamiltonian
is a $2 \times 2$ matrix,
\begin{align}
  H
\approx
\begin{pmatrix}
  H_{11} & \delta m \\
  \delta m & H_{22}
\end{pmatrix},
\label{eq:H}
\end{align}
in the $(n, \bar n)$ basis. The diagonal elements 
$H_{11}$ and $H_{22}$ are the neutron and antineutron
energies \cite{Phillips:2014fgb}.
The Majorana mass $\delta m$ that violates the
baryon number by 2
will induce the $n$-$\bar n$ mixing,
$\tan 2 \theta = 2 \delta m / (H_{22} - H_{11})$,
\begin{align}
\begin{pmatrix}
  n_1 \\
  n_2
\end{pmatrix}
=
  U
\begin{pmatrix}
  n \\
  \bar n
\end{pmatrix},
\quad
  U
=
  \begin{pmatrix}
      \cos \theta & \sin \theta \\
      -\sin \theta & \cos \theta 
  \end{pmatrix},
\label{eq:U}
\end{align}
where $n_1$ and $n_2$ are the two mass eigen-fields.
Typically, the mixing angle $\theta$ is highly suppressed
by the tiny $\delta m$. For simplicity, one
may denote the sine and cosine of the mixing
angle as $(s, c) \equiv (\sin \theta, \cos \theta)$.
The energy difference between
the mass eigenstates is simply denoted as
$\Delta E \equiv \sqrt{(H_{11} - H_{22})^2 + 4 \delta m^2}$
\cite{Phillips:2014fgb}.

The oscillation probability takes the form as,
\begin{align}
  P_{n \bar n}
& =
  \frac 
  { c^2 s^2 (1 - f_1)^2
+ c^2 s^2 (1 - f_2)^2}
  { c^2 (1 - f_1)^2 + s^2 (1 - f_2)^2 }
\nonumber
\\
& -
  \frac{2 (1 - f_1) (1 - f_2)
  c^2 s^2 
  \cos (\Delta E t)}{ c^2 (1 - f_1)^2 + s^2 (1 - f_2)^2 },
\label{eq:Pnnbar}
\end{align}
where $f_1$ and $f_2$ are the phase space distribution
functions of the two mass eigenstates. Note that the
phase space distribution function appears
in the denominator as $1 - f_i$.
In a dense environment with degeneracy, the phase
space distribution function $f_i$ can approach 1 below
the Fermi surface. It is then possible for the oscillation
probabilities to receive enhancement from degeneracy.

With a tiny mixing angle $\theta$
\cite{Phillips:2014fgb}, $s \ll 1$,
the mass eigenstate $n_1$ is approximately the neutron
state $n$ that dominates the matter inside NS.
Consequently, both $1 - f_1$ and $f_2$ are negligibly
small. Between the two tiny terms $s$ and $1 - f_1$ in the
denominator, the case $s \gg 1 - f_1$ would lead to a large
$P_{n \bar n} \approx c^2$ which means degenerate
neutrons would instantly transit to antineutrons. The
situation $s \sim 1 - f_1$ has also the same conclusion.
Since neutron stars are readily observed and hence should be
stable, only the remaining case $s \ll 1 - f_1$ can happen.
This already indicates that the degenerate $n \bar n$
oscillation in NS can give a very much enhanced
constraint.

\begin{figure}[t]
\centering
\includegraphics[width=\linewidth]{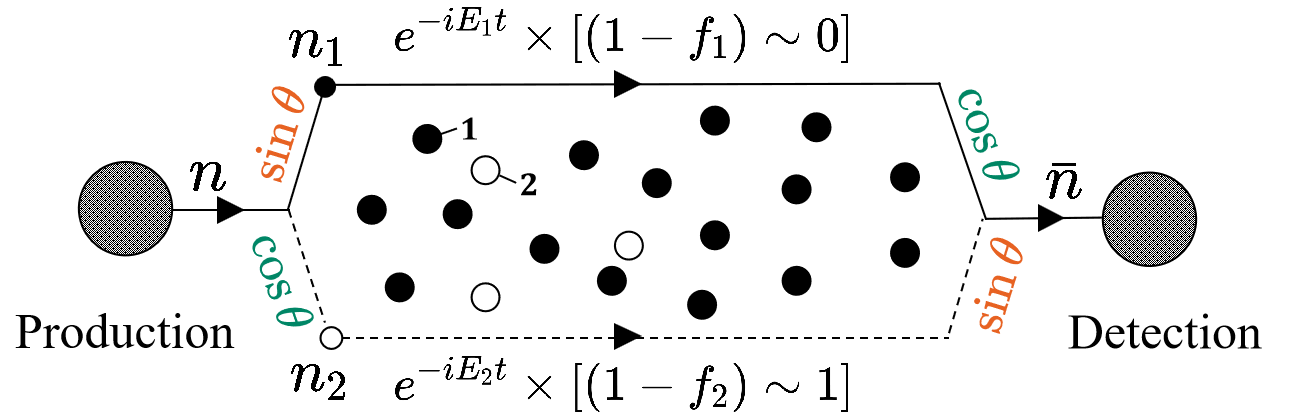}
\caption{Illustration of the degenerate $n$-$\bar n$ oscillation.}
\label{fig:animation}
\end{figure}
In the limit of $s \ll 1 - f_1 \ll 1$
and $f_2 \ll 1$, the transition rate from neutron to
antineutron becomes
\begin{align}
  P_{n \bar n}
\approx
  \frac{s^2}{(1-f_1)^2}
\equiv
  s^2_{\rm eff},
\qquad
  P_{n n}
\approx
  c^2.
\label{eq:P_n-nbar}
\end{align}
As shown in \gfig{fig:animation},
although the neutron state ($n$) is mainly
composed of $n_1$, its coefficient $\cos \theta$
receives extra suppression $1 - f_1$ due to
Pauli blocking. So the relative fraction
of $n_2$ and subsequently the chance for
the antineutron ($\bar n$) to appear
is enhanced. It is much easier
for neutrons in a degenerate neutron environment
to transit to antineutrons without Pauli blocking.
Switching
neutron with antineutron and hence $f_1$ with $f_2$,
the reversed transition probability $P_{\bar n n}
\approx s^2$ is not enhanced. The bidirectional
oscillation has imbalanced rates
to drive the overall transition from the degenerate
neutrons to the non-degenerate antineutrons.

\vspace{1em}

{\bf Standing Fraction of Antineutron in Neutron Star} --
Another important feature is that the time dependence
of the $n \bar n$ oscillation probability
$P_{n \bar n}$, which is
$2 c^2 s^2 \sin^2 (\Delta E t / 2)$ for the vacuum
case, is now suppressed by extra $1-f_1$.
In other words, the
transition happens instantaneously.
Further more, the environment is full of neutrons such
that no particle can freely propagate for sizable
distance. It is unavoidable to consider the zero
time/distance limit.

In an extremely degenerate environment,
neutrons transit into 
antineutrons so rapidly that the ratio between
their distribution functions $f_{\bar n} / f_n$ stays the same,
\begin{align}
  R(\bm p) \equiv \frac{ f_{\bar n} }{ f_n }
=
  \frac {P_{n \bar n}}
        {P_{n n}}
\approx
  \frac{\tan^2 \theta}{[1-f_1(\bm p)]^2},
\label{eq:R}
\end{align}
which actually reduces to the effective mixing $s^2_{\rm eff}$.
With momentum dependence inherited from $f_1(\bm p)$,
$s^2_{\rm eff}$ does not have a fixed value. Even directly
taking the zero time limit of \geqn{eq:Pnnbar},
$P_{n \bar n}(t = 0) \approx s^2 (f_1 - f_2)^2 / (1 - f_1)^2$
where the approximation only applies for the denominator,
\geqn{eq:R} can be readily reproduced with $(f_1 - f_2)^2
\approx 1$. In other words, the production process is
already affected by the degenerate environment as pointed
out earlier.
The neutron transition to antineutron
can instantly happen and remain the same antineutron
fraction over time.

Being in thermal equilibrium as maintained inside NS,
neutrons follow the Fermi-Dirac distribution,
$f_1(\bm p) = 1 / [e^{(\varepsilon_n(\bm p)-\mu)/T}+1]$
where $\varepsilon_n(\bm p) = \sqrt{\bm p^2+m^2}$
is the neutron energy with momentum $\bm p$
and mass $m$. With a very low internal temperature
$T \sim 7$\,keV at the age of $10^5$ years
\cite{Potekhin:2020ttj}, the corresponding neutron
Fermi momentum $k_F=(3 \pi^2 n_n)^{1/3}$ and
chemical potential $\mu =\sqrt{k_F^2+m^2}$
can be readily derived from the neutron density $n_n$
assuming full degeneracy.
We consider a typical NS 
with mass $M=1.4 M_\odot$ (solar mass $M_\odot$) 
and radius $R_n=10$\,km \cite{Lattimer:2000nx}.
To match the corresponding high density
$n_n \approx 4 \times 10^{38}$ cm$^{-3}$ \cite{Potekhin:2015qsa},
which is also assumed to be constant inside NS,
the difference of chemical potential $\mu$ from
the neutron mass $m$,
$\mu -m \approx 101.6$\,MeV, is much larger than
temperature $T$ \cite{Brandes:2023bob}.
It is fully consistent
to say that the distribution function is very
close to 1 below the Fermi surface,
$1 - f_1(\bm p) \approx e^{[\varepsilon_n(\bm p)-\mu]/T} \ll 1$.
Although neutrons can form superfluid, it can
only happen around the Fermi surface with a gap
$\sim 1$\,MeV \cite{Haskell:2017lkl} much smaller than
the chemical potential $\mu - m$ and Fermi momentum.
In addition, $f_1$ deviates from 1 in this region
and hence the $n \bar n$ oscillation would not
experience degeneracy enhancement.

Since the first mass eigenstate is mainly neutron,
$n_1 = c\,n + s\,\bar n$ according to \geqn{eq:U},
they share roughly the same phase space distribution,
$f_n \approx f_1$.
The standing fraction of antineutron as shown in \geqn{eq:R}
determines the antineutron phase space distribution,
$f_{\bar n}(\bm p) = R(\bm p) f_n(\bm p) \approx s^2 f_1(\bm p) / [1 - f_1(\bm p)]$ and the antineutron number density
$n_{\bar n} = \int \frac{g_s}{(2\pi)^3} \mathrm d^3 \bm p f_{\bar n}(\bm p)$
can be analytically obtained,
\begin{equation}
  n_{\bar n}
=
\frac{m^2 T}{2 \pi^2}
e^{\frac \mu T}
\left[
  2 K_{2} 
  \left(\frac m T \right)
+ e^{\frac \mu T} K_{2} \left(\frac {2m} T \right)
\right] s^2 \,,
\label{eq:rhonbar}
\end{equation}
where $K_\alpha(x)$ is the modified Bessel 
function of the second kind. For convenience,
let us define $R \equiv n_{\bar n} / n_n$ that
has no momentum dependence.

\begin{figure}[t]
\centering
\includegraphics[width=1.1\linewidth]{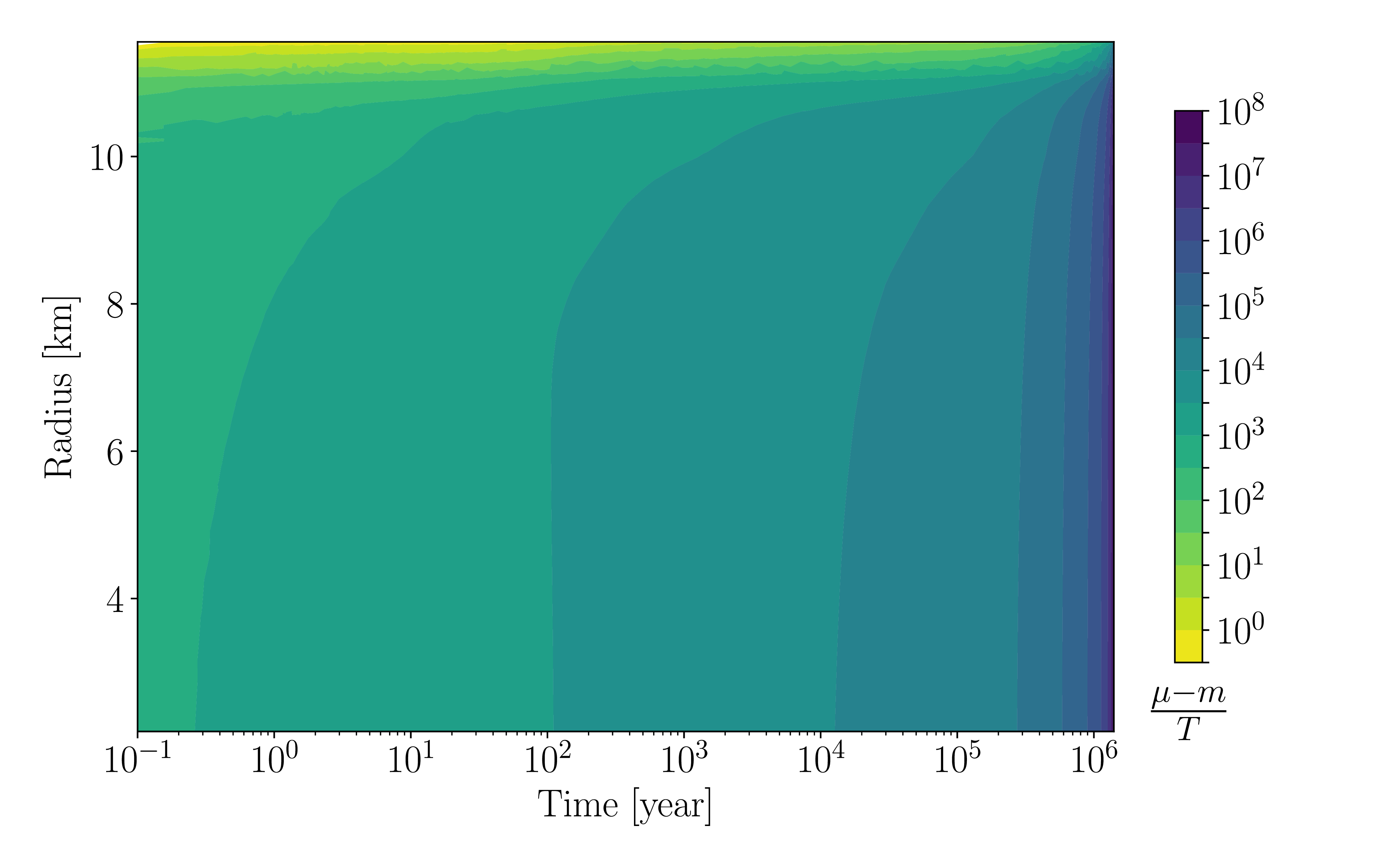}
\caption{The degeneracy parameter $(\mu - m) / T$ of neutron matter
inside NS without considering $n\bar{n}$ annihilation} as function of radius and cooling time.
\label{fig:muT}
\end{figure}

For comparison, the non-degenerate approach gives
an average antineutron fraction
$R' = 2 c^2 s^2$ by neglecting the oscillating part
\cite{Buccella:1987pv}.
Between the degenerate and non-degenerate cases, the difference
in the antineutron fraction
$R / R' \equiv n_{\bar n} / n'_{\bar n}$
can reach $\mathcal O(10^3)$ orders with
$(\mu-m) / T \sim 10^4$ which is a huge enhancement
$e^{(\mu - m) / T}$ due to
degeneracy. As shown in \gfig{fig:muT},
the neutron degeneracy parameter $(\mu - m) / T$
decreases with radius but increases with time according to
\texttt{NSCool}
\cite{Page:2004fy, NSCool program}.

\vspace{1em}

{\bf Neutron Burning and Neutron Star Heating} --
A standing fraction of antineutron
in the dense neutron environment as indicated by
\geqn{eq:R} would lead to constant burning of neutron
matter by neutron-antineutron annihilation.
In a single annihilation, the total nucleon number
$N = N_n + N_{\bar n}$ decreases by two,
$  \mathrm d N
=
- 2 \langle \sigma v \rangle
  n_n n_{\bar n} \mathrm d t \mathrm d V
$,
where $\langle \sigma v \rangle$ is the thermally averaged
cross section $\sigma$ times the relative velocity $v$.
Replacing the antineutron
number density $n_{\bar n} = R n_n$ gives,
\begin{align}
  \frac{\mathrm d N}{\mathrm d t}
=
- \frac {2 R} {1 + R}
  n_n \braket{\sigma v} N
\equiv
- \Gamma N,
\end{align}
where $\Gamma \equiv 2 R n_n \langle \sigma v \rangle / (1 + R)$ is the neutron burning rate.

Suppose the neutron burning does not consume a significant
fraction of the NS mass, the neutron density $n_n$
and the annihilation rate $\Gamma$ can then be treated
as constants if the temperature is also a constant.
Then the total nucleon number decays
as, $N (t) = N (0) e^{- \Gamma t}$. With
$n_n = 10^{38}\,\rm{cm^{-3}}$ and
$\braket{\sigma v} \approx 10^{-15}\,\rm{cm^3/s}$
\cite{Buccella:1987pv},
the neutron burning rate can reach 
$\Gamma \approx 2 R \times 10^{23}\,{\rm{s^{-1}}}$
where we have taken the approximation $R \ll 1$.
For NS to exist for more than one million years,
$\Gamma \lesssim 3 \times 10^{-14}$\,s$^{-1}$, which
would require a lose constraint $R \lesssim 10^{-37}$. The actual
constraint is even much stronger since the temperature
descreases very fast with time and the resulting
degeneracy parameter $(\mu - m) / T$ can reach above
$10^5 \sim 10^6$ \cite{Bertoni:2013bsa} as shown in \gfig{fig:muT}.

The annihilation
products are mainly mesons such as $\pi^0$ and $\pi^\pm$.
While the neutral $\pi^0$ finally decays into two photons
to deposit all its energy as entropy in the NS,
the charged $\pi^\pm$ produce both charged leptons (muon
$\mu^\pm$ and electron/positron $e^\mp$) and neutrinos,
$\pi^\pm \rightarrow \mu^\pm + \nu$ and
$\mu^\pm \rightarrow e^\pm + 2 \nu$. The energies taken
away by neutrinos are actually quite limited since the
energetic pions from the neutron-antineutron annihilation
first lose energy in dense material and then decay at
rest. Similar thing happens for muons. So the neutrino
energy spectrum from pion and muon decays are well defined.
On average, neutrinos take way $\bar E_\nu = 50.4$\,MeV
energy \cite{Liu:2021xzc}. The other energies,
$(m_n - \bar E_\nu) \times d N / d t$,
deposit to heat the NS and affect its cooling curve.
%

\begin{figure}[t]
\centering
\includegraphics[width=0.99\linewidth]{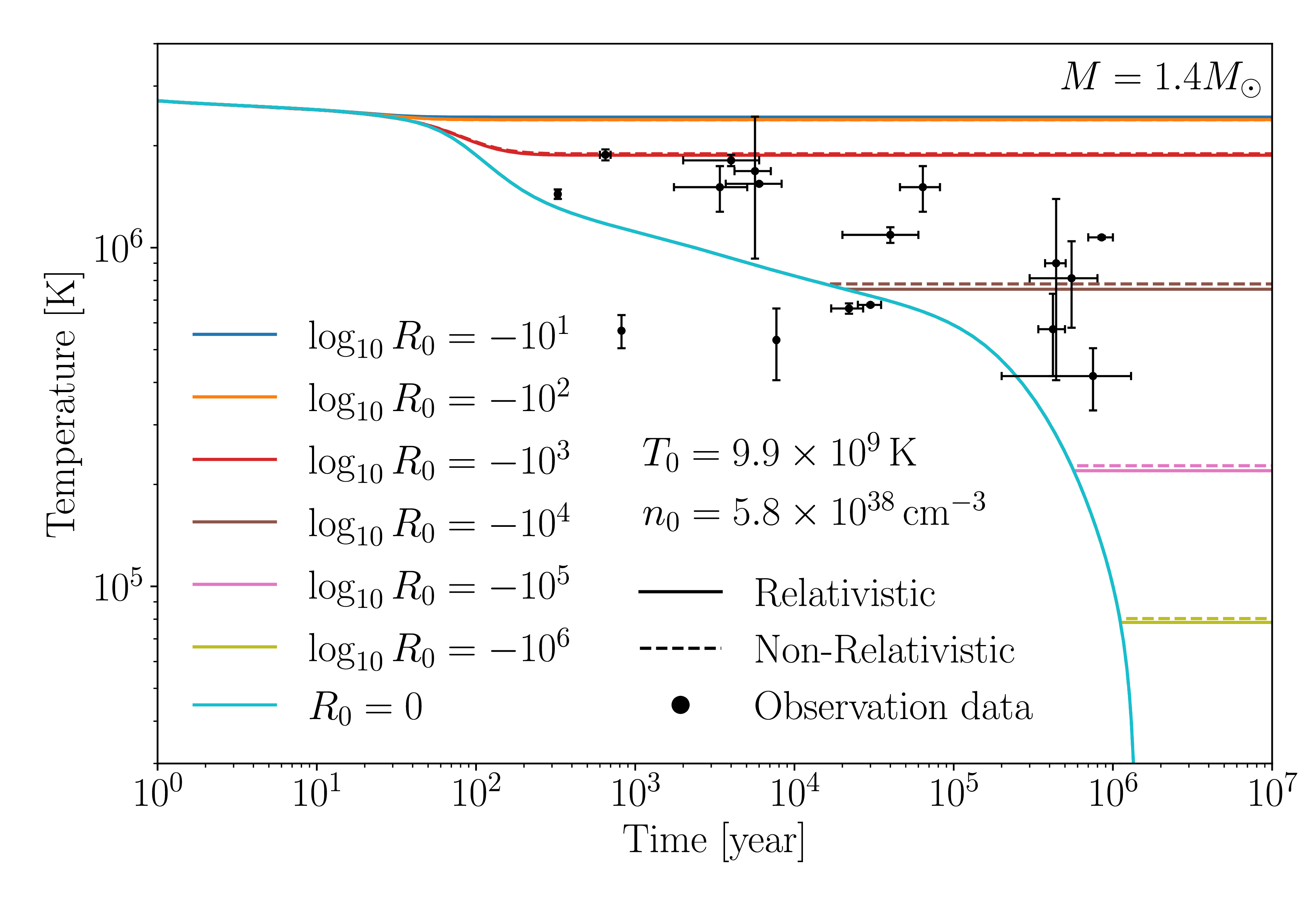}
\caption{The red-shifted NS surface temperature data and cooling curves
for both relativistic (solid) and non-relativistic (dashed)
dispersion relations. Instead of the mixing angle $s^2$,
the input parameter is taken as the antineutron density ratio
$R_0 \equiv R(n_0, T_0)$ with $T_0=9.9\times10^9$\,K at the
beginning of the NS evolution and
$n_0 = 5.8 \times 10^{38}$\,cm$^{-3}$ around the NS center.}
\label{fig:nnosc}
\end{figure}

We implement the heating due to $n \bar n$
oscillation in \texttt{NSCool} \cite{Page:2004fy, NSCool program}
to simulate the NS temperature evolution.
As shown in \gfig{fig:nnosc}, the heating effect
by $n \bar n$ oscillation will prevent NS from
cooling at some point such that the cooling curve
becomes flat. To be consistent with the observed data
points \cite{Potekhin:2020ttj}, the initial parameter $R_0$ should
not exceed $\log_{10} R_0 \sim - \mathcal O(10^4)$. With
$(\mu - m) / T$ reaching $10^6$ \cite{Bertoni:2013bsa} around
$10^5$ years, the physical antineutron fraction
$R \approx (T / T_0)^{3/2} e^{2 (\mu - m) / T - 2 (\mu_0 - m) / T_0} R_0$
is enhanced
to $R \lesssim 10^{-43}$. Correspondingly, the heat release
$2 R n^2_n \langle \sigma v \rangle V m$
in the NS core with a radius of 2\,km can reach
$10^{24}$\,W to far exceed the cooling power
$4 \pi R^2 \sigma T^4 \sim 10^{21}$\,W at the NS surface.
When converted to the mixing angle, the constraint
can reach $s^2 \lesssim 10^{-\mathcal O(10^4)} $.

The mixing angle can be expressed approximately as
$s \sim \delta m / \Delta H$ where $\Delta H \equiv H_{22} - H_{11}$ is the
energy difference between neutron and antineutron
in the range from 1\,MeV \cite{Buccella:1987pv, McKeen:2018xwc} to 1\,GeV
\cite{Berezhiani:2018zvs}. On the other hand, in some GUT models the cross
mass term $\delta m$ is characterized by an effective
mass scale $M_X$, $\delta m \sim \Lambda_{QCD}^6 / M_X^5$
where $\Lambda_{\mathrm{QCD}} \simeq 180$\,MeV
\cite{Phillips:2014fgb}. The enhanced constraint
from NS heating can reach $M_X > 10^{\mathcal O(1000)}$\,GeV
which is far beyond the Planck scale $\sim 10^{19}$\,GeV.

\vspace{1em}

{\bf Summary and Conclusions} --
We establish the formalism of degenerate oscillation
in the QFT language with three different approaches
of second quantization, thermal QFT, and kinetics
theory. Although it can also apply for the bosonic
case, we take the neutron-antineutron
oscillation in NS for a concrete illustration.
Our result shows that this degeneracy
effect can significantly change the neutron oscillation
behavior. Especially, the probed GUT scale can go far beyond
the Planck scale which means that those GUT models
with $n \bar n$ oscillation are essentially excluded by
the stability of NS. Otherwise, NS may explode 
when the degeneracy parameter $(\mu - m) / T$ significantly
increases with much lower internal temperature
towards the end of NS cooling curve. Similar conclusion
should also apply for the neutron-mirror-neutron oscillation
scenarios with cooling effect.

\section*{Acknowledgements}
The authors would like to thank Bo-Wen Ouyang, Pedro Pasquini,
Liang Tan, Jiaming Zheng, 
Albert Zhou, and Wei Zuo for useful discussions.
SFG is particularly grateful to Luke Johns, Georg Raffelt, Guenter Sigl,
and Maria Cristina Volpe for useful comments and discussions.
The authors are supported by the National Natural Science
Foundation of China (12375101, 12090060 and 12090064) and the SJTU Double First
Class start-up fund (WF220442604).
SFG is also an affiliate member of Kavli IPMU, University of Tokyo.
XYF, ZYG, and QHW are supported by the Zhiyuan Future Scholar
program (ZIRC2024-10) of the SJTU Zhiyuan College.

\end{document}